\documentclass[aps,prd,10pt,superscriptaddress,preprint,onecolumn,nofootinbib]{revtex4-1}
\usepackage{graphicx, epsfig} 
\usepackage{amsmath,amssymb,amsfonts,dsfont,mathrsfs,amsthm,mathtools}
\usepackage{bm} 
\usepackage{color}
\usepackage[usenames]{xcolor}
\usepackage{hyperref}
\usepackage{siunitx}
\hypersetup{colorlinks=true,urlcolor=blue,linkcolor=magenta,citecolor=blue,filecolor=blue}
\usepackage[normalem]{ulem}
\usepackage{array}
\usepackage{hyperref}
\usepackage{booktabs}
\usepackage{blindtext}

\newcommand{\diff}[1]{\text{d}#1}

\newcommand{\Lag}{\mathscr{L}}

\begin{document}

\title{Conformal renormalization of scalar-tensor theories}

\author{Giorgos Anastasiou}
\email{ganastasiou@unap.cl}
\affiliation{Instituto de Ciencias Exactas y Naturales, Universidad Arturo Prat, Playa Brava 3256, 1111346, Iquique, Chile}
\affiliation{Facultad de Ciencias, Universidad Arturo Prat, Avenida Arturo Prat Chac\'on 2120, 1110939, Iquique, Chile}

\author{Ignacio J. Araya}
\email{ignaraya@unap.cl}
\affiliation{Instituto de Ciencias Exactas y Naturales, Universidad Arturo Prat, Playa Brava 3256, 1111346, Iquique, Chile}
\affiliation{Facultad de Ciencias, Universidad Arturo Prat, Avenida Arturo Prat Chac\'on 2120, 1110939, Iquique, Chile}

\author{Mairym Busnego-Barrientos}
\email{mbusnego@estudiantesunap.cl}
\affiliation{Instituto de Ciencias Exactas y Naturales, Universidad Arturo Prat, Playa Brava 3256, 1111346, Iquique, Chile}
\affiliation{Facultad de Ciencias, Universidad Arturo Prat, Avenida Arturo Prat Chac\'on 2120, 1110939, Iquique, Chile}

\author{Crist\'obal Corral}
\email{crcorral@unap.cl}
\affiliation{Instituto de Ciencias Exactas y Naturales, Universidad Arturo Prat, Playa Brava 3256, 1111346, Iquique, Chile}
\affiliation{Facultad de Ciencias, Universidad Arturo Prat, Avenida Arturo Prat Chac\'on 2120, 1110939, Iquique, Chile}

\author{Nelson Merino}
\email{nemerino@unap.cl}
\affiliation{Instituto de Ciencias Exactas y Naturales, Universidad Arturo Prat, Playa Brava 3256, 1111346, Iquique, Chile}
\affiliation{Facultad de Ciencias, Universidad Arturo Prat, Avenida Arturo Prat Chac\'on 2120, 1110939, Iquique, Chile}

\begin{abstract}
We study a conformally coupled scalar-tensor theory with a quartic potential possessing local conformal symmetry up to a boundary term. We show that requiring the restoration of the full local conformal symmetry fixes the counterterms that render the on-shell action finite. The building block of the resulting action is a conformally covariant tensor which is constructed out of the metric and the scalar field and it has the same conformal weight as the Weyl tensor. This allows us to obtain the counterterms for the scalar-tensor sector in a closed form. The finiteness of the conformally complete version of the action is suggestive on the validity of the \emph{Conformal Renormalization} prescription.  We extend this theory by adding the Conformal Gravity action and also the Einstein-AdS action written in MacDowell-Mansouri form. Even though the latter breaks the conformal symmetry, we find that the action is still renormalized provided a suitable falloff of the scalar field when considering asymptotically locally anti-de Sitter solutions. Black hole solutions in these theories are studied, for which the Hawking temperature and the partition function to first order in the saddle-point approximation are calculated, providing a concrete example of this renormalization scheme.
\end{abstract}

\maketitle

\section{Introduction}

Renormalization of gravitational theories that admit asymptotically locally anti-de Sitter (AlAdS) solutions has become a crucial ingredient in black hole thermodynamics and in the anti-de Sitter/conformal field theory (AdS/CFT) correspondence. The standard prescription ---dubbed holographic renormalization (HR)--- consists in adding the Gibbons-Hawking-York (GHY) term in order to fix the Dirichlet variational principle for the induced metric in the radial foliation and then introducing intrinsic boundary counterterms to cancel divergences that appear when evaluating the action on solutions with AlAdS behavior. This renders the Euclidean on-shell action and asymptotic charges finite~\cite{Henningson:1998gx,Balasubramanian:1999re,Chamblin:1999tk,Emparan:1999pm,Chamblin:1999hg,Nojiri:1999mh,deHaro:2000vlm,Bianchi:2001kw,Skenderis:2002wp}, which allows one to define the generating functional for correlators of the dual CFT and to obtain the holographic data of said theory~\cite{Gubser:1998bc,Witten:1998qj}. In particular, in the saddle-point approximation of the AdS/CFT correspondence, the
Euclidean on-shell action of the gravity theory is identified with the generating functional for connected correlators of the CFT, where the values of the fields at the conformal boundary are identified with the holographic sources of the corresponding CFT operators. Then, the variational principle ensures that arbitrary variations of the generating functional are expressed as total variations of the sources, such that the correlators are directly obtained through functional derivatives thereof. 

An interesting observation, which was made in Ref.~\cite{Papadimitriou:2005ii}, is that imposing the Dirichlet condition for the induced metric at the AdS boundary is ill-defined due to the divergent volume element of AdS space. Furthermore, fixing the Dirichlet condition for the holographic source asymptotically close to the boundary does not require fixing the Dirichlet condition for the intrinsic metric, as both the extrinsic and intrinsic curvatures admit an expansion in the holographic source in terms of the Fefferman-Graham (FG) expansion~\cite{AST_1985__S131__95_0,Graham:1999jg}. This lead to the development of the Kounterterms prescription in Ref.~\cite{Olea:2006vd}, where the renormalization is achieved by the addition of a suitable boundary term, which depends on both extrinsic and intrinsic curvatures of the boundary in a closed form. For even bulk dimensions, the Kounterterm is the Chern form, which is the boundary term that appears in the Euler theorem. Then, the renormalization can also be achieved by adding the Euler density in the bulk with a fixed coupling, such that it cancels the divergence of the maximally-symmetric configuration (global AdS). In the case of odd bulk dimensions, there is a similar procedure where the boundary term is the contact term of the transgression form of the AdS group~\cite{Mora:2006ka}. In that case, the second gauge connection describes a product manifold that shares the same boundary as the dynamical manifold.

As shown in Ref.~\cite{Miskovic:2009bm}, the 4D Kounterterm-renormalized Einstein-AdS action can be written in a McDowell-Mansouri form for the AdS group~\cite{Stelle:1976gc,MacDowell:1977jt}. In the case of Einstein-AdS spacetimes, the latter can be written in terms of the Weyl tensor squared and the on-shell action becomes that of Conformal Gravity (CG), which is the unique local conformal invariant in 4D. This rewriting of the Einstein theory by embedding it in CG is consistent at the level of the equation of motion (EOM), as all Einstein spacetimes belong to the solution space of the theory, whose EOM is given by the Bach-flat condition. Since CG contains an Einstein sector, the corresponding action can be explicitly separated into a MacDowell-Mansouri part plus terms which vanish for Einstein spaces~\cite{Anastasiou:2016jix}.

For AlAdS manifolds with weakened AdS asymptotics, it was shown that the Weyl-squared CG action is finite off-shell, even for non-Bach-flat spacetimes~\cite{Grumiller:2013mxa}. Therefore, the finiteness of the McDowell-Mansouri action for Einstein-AdS gravity follows immediately, as the two actions are equivalent for Einstein spaces. This was the first example where the embedding of a gravity theory into another one with bulk local conformal invariance allowed to obtain the renormalized form of the action. Later, in Ref.~\cite{Anastasiou:2020mik}, the same procedure was generalized for Einstein-AdS gravity in 6D by embedding it into the unique CG action in 6D which admits Einstein spaces as solutions, constructed in Ref.~\cite{Lu:2011zk}. It is important to emphasize that, even though the embedding of Einstein-AdS into CG in 4D gives the same action principle as the Kounterterms, this is not true in 6D. As discussed in Ref.~\cite{Anastasiou:2020zwc}, in the 6D case, the topologically renormalized action cancels all divergences only for AlAdS spaces with conformally flat boundary. In the generic case, the latter prescription fails to cancel a boundary divergence which depends of the Weyl-squared of the boundary manifold. However, the embedding in the 6D CG theory correctly reproduces all the terms required for achieving the renormalization, such that the obtained action is fully equivalent to the one given by HR, up to the normalizable order. Thus, it is this prescription that gives the correct renormalization, generalizing the Kounterterms beyond the requirement of the conformal flatness of the boundary.

Beyond purely metric theories, renormalization approaches have been considered for cases with additional degrees of freedom, e.g. scalar fields. Indeed, in the context of HR, counterterms for AlAdS spaces in scalar-tensor theories have been discussed in Refs.~\cite{Nojiri:1998dh,Padilla:2012ze,Caldarelli:2016nni,Liu:2017kml,Li:2018rgn,Agurto-Sepulveda:2022vvf}. Therefore, a natural question to ask is whether or not the use of local conformal symmetry in the bulk to determine the renormalization terms can be generalized for these theories. Here, we address this issue and construct renormalized gravity actions possessing a conformally coupled scalar-tensor sector, whose solutions have been studied in the literature. Therefore, this constitutes the first application of the \emph{Conformal Renormalization} idea to scalar-tensor theories of gravity.

The paper is organized as follows. In Sec.~\ref{Sec:CG}, we review the purely metric formulation of CG in four dimensions and show its finiteness for AlAdS spaces as well as its equivalence with renormalized Einstein-AdS theory when evaluated at Einstein spaces. In Sec.~\ref{Sec:CCS}, we obtain the conformal completion of a conformally coupled scalar-tensor theory, which allows us to read the counterterms for the AlAdS sector explicitly. Section~\ref{Sec:F2Sigma2} is devoted to present the renormalized Einstein-AdS gravity in MacDowell-Mansouri form conformally coupled to scalar fields. We show that, even though the latter is not conformal invariant, the counterterms still render the action finite for a suitable falloff of the scalar field; we compute its value explicitly for the solution obtained in Refs.~\cite{Martinez:2002ru,Martinez:2005di}. In Sec.~\ref{Sec:CGscalars}, we prove the renormalization of CG with conformally coupled scalar fields for all Bach-flat solutions and we compute explicitly the renormalized action for analytic black-hole configurations with stealth scalar fields. Finally, in Sec.~\ref{Sec:discussion} we present a summary and discussion about the main results.     

\section{Conformal Gravity\label{Sec:CG}}

In four dimensions, CG is an interesting theory constructed solely in terms of the squared Weyl tensor. It has been studied as an ultraviolet completion of General Relativity~\cite{Adler:1982ri}, as a counterterm in holographic renormalization~\cite{Sen:2012fc}, and it appears as a possible explanation for the flat galaxy rotation curves~\cite{Mannheim:1988dj}. Different supersymmetric extensions of CG have been studied in Refs.~\cite{Kaku:1977pa,Kaku:1978ea,Kaku:1978nz,deWit:1980lyi,Bergshoeff:1980is,Fradkin:1985am,Liu:1998bu,Butter:2009cp,Ferrara:2018wqd,Andrianopoli:2014aqa,DAuria:2021dth} alongside their holographic properties~\cite{Andrianopoli:2020zbl}. Additionally, it arises in the context of the twistor formulation of string theory~\cite{Berkovits:2004jj}. Its action represents the only four-dimensional functional constructed uniquely in terms of the metric that remains invariant under local Weyl rescalings $g_{\mu\nu}\to\tilde{g}_{\mu\nu} = e^{2\sigma(x)}g_{\mu\nu}$. As a consequence, the theory involves four-derivative terms which makes it pathological due to the presence of ghosts. However, since higher-derivative theories of gravity have better renormalizability properties than Einstein gravity~\cite{Capper:1975ig,Stelle:1976gc,Julve:1978xn,Fradkin:1981iu}, they are considered useful toy models for quantum gravity. 

The space of solutions of CG contains all Einstein spaces. Indeed, the CG action becomes equal to that of renormalized Einstein-AdS gravity when evaluated at Einstein spaces~\cite{Anastasiou:2016jix}. In the case of AlAdS manifolds, the Einstein condition can be implemented, up to the normalizable order, by imposing Neumann boundary conditions on the FG expansion~\cite{Maldacena:2011mk}. Additionally, this theory has the remarkable property of being finite and possessing a well-posed variational principle for AlAdS spacetimes~\cite{Grumiller:2013mxa}. Specifically, the dynamics of Conformal Gravity is dictated by the action principle
\begin{equation}
I_{\rm CG} = \alpha_{\rm CG} \int\diff{^4x}\sqrt{|g|} \;W^{\alpha \beta}_{\mu \nu} W^{\mu \nu}_{\alpha \beta} \,,
\label{ICGaction}
\end{equation}
where $\alpha_{\rm CG}$ is a dimensionless coupling constant, $g=\det g_{\mu\nu}$ is the metric determinant, while
\begin{equation}
W_{\mu \nu}^{\alpha \beta}= R_{\mu \nu}^{\alpha \beta} -4S^{[\alpha}_{[\mu} \delta^{\beta]}_{\nu]} \;\;\;\;\; \mbox{and} \;\;\;\;\; S_{\mu\nu} = \frac{1}{2} \left(R_{\mu \nu} - \frac{1}{6} g_{\mu \nu} R\right) \,,
\label{weyltensor}
\end{equation}
are the Weyl and Schouten tensor, respectively. Here, Greek indices indicate the bulk coordinate patch. The four-order field equations are obtained by performing arbitrary variations of Eq.~\eqref{ICGaction} with respect to the metric, giving $B_{\mu\nu}=0$, where
\begin{align}
B_{\mu \nu} = -4 \left(\nabla^{\lambda} C_{\mu \nu \lambda} + S^{\lambda \sigma} W_{\mu \lambda \nu \sigma } \right)\;\;\;\;\; \mbox{and} \;\;\;\;\;
C_{\mu \nu \lambda} = \nabla_{\lambda} S_{\mu \nu} - \nabla_{\nu} S_{\mu \lambda} \,, 
\end{align}
are the Bach and Cotton tensors, respectively. Therefore, the solution space of the theory corresponds to Bach-flat spacetimes. Moreover, notice that Einstein spaces satisfy this condition automatically. Then, all Einstein spaces are Bach flat, even though the converse is not necessarily true. On the other hand, even though some Bach-flat spaces are conformally Einstein, there are examples where this condition is not satisfied and they cannot be related to solutions in Einstein gravity by performing a conformal transformation~\cite{Liu:2013fna,Dunajski:2013zta}. 

One of the most interesting features of CG is the finiteness of the action when evaluated at AlAdS spacetimes. Namely, the usual divergences that arise in the gravitational action due to the infinite volume of AdS spaces are absent in the case of CG and no additional counterterms are needed. Indeed, as shown in Ref.~\cite{Grumiller:2013mxa} for weakened asymptotically AdS boundary conditions, both the quasilocal stress tensor and the partially massless response function are finite; these are the corresponding currents coupled to the massless and the massive gravitons, respectively. These independent currents are the ones that in AdS/CFT are identified with the CFT operators whose holographic sources are then the two lowest order coefficients of the FG expansion of the metric about the AdS boundary.

The asymptotic behavior of the CG action can be also studied by considering power-counting arguments. In particular, we consider the generic AlAdS conditions which in the FG gauge obtain the form
\begin{align}
ds^{2}&= \frac{\ell^2}{z^2} \left(dz^{2}+ \mathcal{G}_{i j} \left(z,x\right) dx^{i} dx^{j} \right)\,, \notag \\
\mathcal{G}_{i j} \left(z,x\right) &= g_{\left(0\right) ij} \left(x\right) + \frac{z}{\ell} g_{\left(1\right) ij} \left(x\right) + \frac{z^2}{\ell^2} g_{\left(2\right) ij} \left(x\right) + \frac{z^3}{\ell^3} g_{\left(3\right) ij} \left(x\right)+ \ldots \,,
\end{align}
where $z$ is the radial coordinate, $\ell$ is the AdS radius, and Latin indices indicate the coordinates at the codimension-1 constant-$z$  hypersurface. Here, $z=0$ denotes the location of the conformal boundary. This structure defines a radial ADM-like foliation that allows us to decompose the Weyl squared in term of the three independent contributions; they are
\begin{equation}
W^{\alpha \beta}_{\mu \nu} W^{\mu \nu}_{\alpha \beta} = W^{ij}_{km} W^{km}_{ij} + 4 W^{iz}_{jz} W^{jz}_{iz} + 4 W^{iz}_{km}W^{km}_{iz} \,.
\end{equation}
Interestingly enough, for the generic AlAdS conditions described above, all the independent components of the Weyl tensor fall-off as $\mathcal{O} \left(z^2\right)$. Thus, the CG Lagrangian falls-off as $\mathcal{O} \left(z^4\right)$, what leads to the action behaving as
\begin{equation}
\int\diff{^4x}\sqrt{|g|}\; W^{\alpha \beta}_{\mu \nu} W^{\mu \nu}_{\alpha \beta} \sim \int \diff{^3x} \int \diff{z}\; \frac{\sqrt{|g_{\left(0\right)}|}}{z^4}\; \mathcal{O} \left(z^4\right) \sim \mathcal{O} \left(z\right) \,.
\end{equation}
The latter indicates that the CG action is free from any IR divergences, in accordance to Ref.~\cite{Grumiller:2013mxa}. This behavior of the CG action makes manifest the relation between bulk conformal symmetry and renormalization, not only for CG, but also for every subsector of the solution space of the theory. Indeed, for Einstein-AdS spacetimes, where $S_{\mu \nu} = -\frac{1}{2\ell^2} g_{\mu \nu}$, the Weyl tensor coincides with the curvature of the torsionless AdS group, $\mathcal{F}^{\alpha \beta}_{\mu \nu}$, given by
\begin{equation}\label{WeylE}
W^{\alpha \beta}_{\left(E\right) \mu \nu} = R^{\alpha \beta}_{\mu \nu} + \frac{1}{\ell^2}  \delta^{\alpha \beta}_{\mu \nu} \equiv \mathcal{F}^{\alpha \beta}_{\mu \nu} \,.
\end{equation}
This relation indicates that the CG action evaluated for Einstein spacetimes reduces to the MacDowell-Mansouri action for the AdS group \cite{MacDowell:1977jt}. The latter corresponds to the topologically renormalized Einstein-AdS action\footnote{Actually, the resulting action in Eq.\eqref{IMM} is shifted by a constant term involving the Euler characteristic of the manifold, which naturally arises from the Kounterterms renormalization scheme and matches the renormalized volume \cite{Anastasiou:2018mfk}.} \cite{Miskovic:2009bm}
\begin{equation}
I_{\rm CG} \left[E\right] = I^{\rm (ren)}_{\rm EAdS}=\frac{\ell^2}{256 \pi G_{N}} \int\diff{^4x}\sqrt{|g|} \delta^{\mu_1\ldots\mu_4}_{\nu_1\ldots\nu_4}\mathcal{F}^{\nu_1\nu_2}_{\mu_1\mu_2}\mathcal{F}^{\nu_3\nu_4}_{\mu_3\mu_4} \,,
\label{IMM}
\end{equation}
which, in turn, is equivalent to holographic renormalization \cite{Anastasiou:2020zwc}.

It is important to note that the action of Eq.~\eqref{IMM} has a well-defined Dirichlet variational principle for the metric at the conformal boundary $g_{\left(0\right)ij}$ \cite{Anastasiou:2019ldc}. This is in accordance with Ref.~\cite{Papadimitriou:2005ii}, which showed that the finiteness and the well-posedness of the variational principle in terms of the holographic sources are related.

Furthermore, Eq.~\eqref{IMM} suggests that the counterterms of Einstein-AdS gravity are dictated by bulk conformal symmetry, which introduces the concept of Conformal Renormalization. Conversely, one could show the finiteness of the MacDowell-Mansouri action~\eqref{IMM} by using the generic off-shell relation between $\mathcal{F}^{\alpha \beta}_{\mu \nu}$ and the Weyl tensor. This is given by
\begin{equation}
W^{\alpha \beta}_{\mu \nu} = \mathcal{F}^{\alpha \beta}_{\mu \nu}- X^{\alpha \beta}_{\mu \nu} \,,
\label{weylfdecomposition}
\end{equation}
\begin{equation}\label{Xtensor}
X^{\alpha \beta}_{\mu \nu} = 2 H^{[\alpha}_{[\mu} \delta^{\beta]}_{\nu]}+ \frac{1}{12} \left(R+\frac{12}{\ell^2} \right) \delta^{\alpha \beta}_{\mu \nu} \,,
\end{equation}
with $H^{\alpha}_{\mu}=R^\alpha_\mu - \tfrac{1}{4}\delta^\alpha_\mu\,R $ being the traceless Ricci tensor. Thus, replacing this relation into Eq.~\eqref{IMM}, one obtains
\begin{equation}\label{IEHren}
I^{\rm (ren)}_{\rm EAdS} = \frac{\ell^2}{256 \pi G_{N}} \int\diff{^4x}\sqrt{|g|} \left[\delta^{\mu_1\ldots\mu_4}_{\nu_1\ldots\nu_4}W^{\nu_1\nu_2}_{\mu_1\mu_2} W^{\nu_3\nu_4}_{\mu_3\mu_4}+ 8 H^{\mu}_{\nu}H^{\nu}_{\mu}+\frac{2}{3} \left(R+\frac{12}{\ell^2}\right)^{2}\right] \,.
\end{equation}
This action reduces exactly to that of Eq.~\eqref{ICGaction} for Einstein spaces, as it can be seen by noticing that the last two terms of Eq.~\eqref{IEHren} vanish identically upon this condition. Therefore, as the CG action is finite in general for weakened AdS asymptotics, one concludes that the obtained expression is renormalized for Einstein-AdS spacetimes.

\subsection{A new route to the renormalized Einstein-AdS action}

In what follows, we provide a different strategy to address Conformal Renormalization. Instead of finding a conformally invariant theory and evaluating it at different sectors of the solutions space, we perform its on-shell conformal completion. This will be the guiding principle for the derivation of the generalization to scalar-tensor theories.

Our starting point is the Einstein-AdS gravity in four dimensions. In this case, the Einstein-Hilbert Lagrangian density with a negative cosmological constant $\Lambda=-3/\ell^2$ reads
\begin{equation}
\Lag_{\rm EH}= \sqrt{|g|} \left(R+\frac{6}{\ell^2}\right) \,.
\end{equation}
The behavior of the Ricci scalar under infinitesimal local Weyl rescalings of the metric, i.e. $\delta_{\sigma} g_{\mu \nu} = 2 \sigma g_{\mu \nu}$, is given by
\begin{equation}
\delta_{\sigma} R = -2 \sigma R - 2\left(D-1\right) \Box \sigma \,.
\end{equation}
Then, the Einstein-Hilbert term transforms according to
\begin{equation}
\delta_{\sigma} \left[\sqrt{|g|} \left(R + \frac{6}{\ell^2} \right)\right]
=2 \sqrt{|g|} \left [\sigma  \left(R +\frac{12}{\ell ^{2}}\right) -3  \nabla ^{\mu } \nabla _{\mu }\sigma \right ] \,. \label{LEH4D}
\end{equation}
In order to conformally complete the Einstein-Hilbert Lagrangian without modifying the field equations, one has to add either a surface or topological terms. Since we are working in four dimensions, we consider that the Einstein-Hilbert action is supplemented by the Gauss-Bonnet term with an arbitrary coupling constant $c_4$, that is,
\begin{equation}\label{LagEHGB}
\Lag_{\rm EH,GB} = \sqrt{|g|} \left(R+\frac{6}{\ell^2} + c_{4} E_{4}\right) \,,
\end{equation}
where $c_4$ is a constant with units of length squared and
\begin{equation}
E_{4} = \dfrac{1}{4} \delta _{\nu _{1} \ldots  \nu _{4}}^{\mu _{1} \ldots  \mu _{4}} R_{\mu _{1} \mu _{2}}^{\nu _{1} \nu _{2}} R_{\mu _{3} \mu _{4}}^{\nu _{3} \nu _{4}} = R^2 - 4R^\mu_\nu R^\nu_\mu + R^{\mu\nu}_{\lambda\rho}R^{\lambda\rho}_{\mu\nu} \,,
\end{equation}
is the Gauss-Bonnet invariant. In four dimensions, the latter does not contribute to the bulk dynamics since its integral is proportional to the sum of the Euler characteristic and the integral of the Chern form in a codimension-one boundary. Nevertheless, it changes the conserved charges and the Euclidean on-shell action in a nontrivial way~\cite{Aros:1999id,Aros:1999kt,Miskovic:2009bm}. Considering its Weyl variation
\begin{equation}
\dfrac{1}{4} \delta _{\sigma } \left(\sqrt{|g|} \;\delta _{\nu _{1} \ldots  \nu _{4}}^{\mu _{1} \ldots \mu _{4}} R_{\mu _{1} \mu _{2}}^{\nu _{1} \nu _{2}} R_{\mu _{3} \mu _{4}}^{\nu _{3} \nu _{4}}\right)
= -16 \delta _{\nu _{1} \nu _{2}}^{\mu _{1} \mu _{2}}  \nabla _{\mu _{2}}\left (\sqrt{|g|} \;S_{\mu _{1}}^{\nu _{1}}  \nabla ^{\nu _{2}}\sigma \right ) \,, \label{weylvarGB}
\end{equation}
and summing up all the contributions, we get
\begin{equation}\delta _{\sigma } \Lag_{\rm EH,GB} 
=2 \sqrt{|g|} \left [\sigma  \left(R +\frac{12}{\ell ^{2}}\right) -3  \Box\sigma  -8 c_{4} \delta _{\nu _{1} \nu _{2}}^{\mu _{1} \mu _{2}} \left (\dfrac{1}{2} C_{\mu _{1} \mu _{2}}^{\nu _{1}}  \nabla ^{\nu _{2}}\sigma  +S_{\mu _{1}}^{\nu _{1}}  \nabla _{\mu _{2}} \nabla ^{\nu _{2}}\sigma \right )\right ]\,.
\end{equation}
As expected, the EH action supplemented by the GB density is not Weyl invariant. However, evaluating at Einstein spaces, the variation becomes 
\begin{equation}
\delta _{\sigma } \Lag_{\rm EH,GB}\vert _{E} =\sqrt{|g|} \left ( -6  \Box\sigma  +\frac{24}{\ell^2} c_{4}  \Box\sigma \right ) \,.
\end{equation}
Demanding on-shell Weyl invariance of the action, the coupling constant is uniquely fixed as $c_{4}=\frac{\ell^2}{4}$. Thus, the corresponding action matches, up to the Euler characteristic of the manifold, the topologically renormalized Einstein-AdS action~\cite{Miskovic:2009bm}. This has recently shown to be equivalent to the HR prescription~\cite{Anastasiou:2020zwc}.
This procedure provides an alternative route to obtain the surface terms that render the action finite.

The generalization of the concept of Conformal Renormalization in six dimension has been presented in Ref.~\cite{Anastasiou:2020mik}. In the following, we will extend this prescription to the case when scalar fields are included.

\section{Conformally coupled scalar fields\label{Sec:CCS}}

The previous analysis indicates that, in the presence of purely metric fields, bulk conformal invariance leads to a finite action when AlAdS spacetimes are considered. In this section, we study whether this relation can be extended in the presence of scalar fields. Our starting point is the conformally coupled scalar-tensor theory with a self-interacting scalar field, whose action is given by
\begin{equation}\label{ccscalar}
I_\phi = \int\diff{^4x}\sqrt{|g|}\left(\frac{1}{12}\phi^{2} R + \frac{1}{2}\nabla^{\mu} \phi \nabla_{\mu} \phi + \nu\phi^{4}\right) \,,
\end{equation}
where $\nu$ is the dimensionless coupling of the quartic scalar potential. This action is quasi-conformal invariant, namely, it transforms as a boundary term under the simultaneous Weyl rescaling of the metric and scalar field given by $g_{\mu\nu}\to\tilde{g}_{\mu\nu} = e^{2\sigma(x)}g_{\mu\nu}$ and $\phi\to\tilde{\phi} = e^{-\sigma(x)}\phi$. Indeed, considering the infinitesimal Weyl rescaling of the fields 
\begin{align}
\delta_{\sigma} g_{\mu \nu} = 2 \sigma g_{\mu \nu} \;\;\;\;\; \mbox{and} \;\;\;\;\; \delta_{\sigma} \phi = - \sigma \phi \,, 
\end{align}
the Eq. \eqref{ccscalar} transforms as
\begin{align}
\delta_{\sigma}I_{\phi} 
&=-\frac{1}{2}\int\diff{^4x}\sqrt{|g|} \nabla_{\mu} \left(\phi^{2} \nabla^{\mu} \sigma \right) \,. \label{Weylvargandphi}
\end{align}
The presence of the total derivative indicates that the action $I_{\phi}$ should be supplemented by a boundary term and/or a topological contribution for restoring the exact local conformal invariance of the theory. 

In order to perform the conformal completion of the last expression, we consider that for a scalar $\Phi$ of arbitrary scaling dimension $\Delta$, the Weyl variation of the Laplacian multiplied with the volume element reads
\begin{equation}
\delta_{\sigma}\left(\sqrt{|g|} \Box \Phi \right) =  \sqrt{|g|} \left[\left(D+\Delta-2\right) \sigma \Box \Phi +\Delta \Phi \Box \sigma + \left(D+2\Delta-2\right) \nabla^{\lambda} \sigma \nabla_{\lambda} \Phi \right] \,.
\end{equation}
Thus, for $\Phi = \phi^2$ with $\Delta=-2$ in four dimensions, we obtain
\begin{equation}
\delta_{\sigma}\left(\sqrt{|g|} \Box \phi^2 \right) = -2 \sqrt{|g|} \nabla_{\mu} \left(\phi^{2} \nabla^{\mu} \sigma \right) \,.
\end{equation}
Therefore, the combination
\begin{equation}
I_{\phi,\rm cc}=I_{\phi} - \frac{1}{4}\int\diff{^4x}\sqrt{|g|} \Box \phi^2 \,,
\end{equation}
is fully invariant under Weyl rescalings of both the metric and the scalar field. Indeed, this action can be equivalently written as
\begin{equation}
I_{\phi,\rm cc}=\int\diff{^4}x\sqrt{|g|}\left(\frac{1}{12}\phi^2 R - \frac{1}{2}\phi\Box\phi + \nu\phi^4\right) \,.
\label{Ibulkconf}
\end{equation}
In the last expression, it becomes manifest that the conformal completion of the non-minimally coupled scalar field action leads to an explicit dependence from the Yamabe operator $\Delta_2$, that reads
\begin{equation}\label{Yamabe}
\Delta_2 = - \Box+ \frac{\left(D-2\right)}{4\left(D-1\right)} R\,.
\end{equation}
This differential operator ---frequently dubbed conformal Laplacian--- is conformally covariant  with a scaling weight $- \frac{D+2}{2}$ when acts on scalars of scaling dimensions $\Delta = - \frac{D-2}{2}$~\cite{Osborn:2015rna,Gover:2002ay}. One may trivially extend this differential operator by adding a scalar with conformal weight $\Delta=-2$. In particular, for the theory of interest, we consider the extension 
\begin{equation}
\tilde{\Delta}_{2} = \Delta_{2}+ c \phi^{2}\,,
\label{Yamabeext}
\end{equation}
where $c$ is a dimensionless arbitrary constant. This allows us to write the action \eqref{Ibulkconf} as
\begin{equation}
I_{\phi,\rm cc}= \frac{1}{2}\int\diff{^4}x\sqrt{|g|} \phi \tilde{\Delta}_{2} \phi \,.
\label{Ibulkconf2}
\end{equation}
Since the Yamabe operator is conformally covariant, the action~\eqref{Ibulkconf2} is explicitly invariant under local rescalings. However, there are configurations which break the symmetry, e.g. constant scalar fields. In that case, the kinetic part of the scalar field vanishes and one is left only with the Einstein-Hilbert part, losing the information of the presence of the scalar. Moreover, it is easy to see that the conformal transformation becomes singular for a constant scalar by starting with a nontrivial scalar configuration such that the action~\eqref{Ibulkconf2} is finite and the spacetime is AlAdS. Then, one can choose the conformal transformation to make the scalar field constant. If the original value of the action  was finite, then, by conformal invariance, it should remain finite at the constant scalar field configuration. However, in that case, the action will diverge as the AdS volume. Therefore, the transformation has to be singular. This implies that $I_{\phi,\rm cc}$ has to be supplemented by the corresponding compensating terms which are necessary to ensure the bulk conformal invariance of all possible field configurations of the theory, including the case of a constant scalar field.

In order to circumvent this issue, it is convenient to introduce the tensor proposed in Ref.~\cite{Oliva:2011np}, namely,
\begin{equation}\label{Stensor}
\mathcal{S}^{\mu\nu}_{\,\lambda\rho}=\phi^2R^{\mu\nu}_{\,\lambda\rho}-4\phi\delta^{[\mu}_{[\lambda}\nabla^{\nu]}\nabla_{\rho]}\phi+8\delta^{[\mu}_{[\lambda}\nabla^{\nu]}\phi\nabla_{\rho]}\phi-\delta^{\mu\nu}_{\lambda\rho}\nabla_\alpha\phi\nabla^\alpha\phi\,,
\end{equation}
which transforms covariantly under Weyl rescalings in four dimensions, i.e.,
\begin{align}
\mathcal{S}^{\mu\nu}_{\lambda\rho}\to \tilde{\mathcal{S}}^{\mu\nu}_{\lambda\rho} &=  e^{-4\sigma(x)}\mathcal{S}^{\mu\nu}_{\lambda\rho}\,. 
\end{align} 
This implies that the tensor~\eqref{Stensor} becomes a convenient building block for constructing conformally invariant scalar-tensor theories of gravity. Indeed, their traces are 
\begin{align}\label{Smunu}
\mathcal{S}^\mu_\nu &\equiv \mathcal{S}^{\mu\lambda}_{\nu\lambda} = \phi^2R^\mu_\nu - \delta^\mu_\nu \phi\Box\phi - 2\phi\nabla^\mu\nabla_\nu\phi + 4\nabla^\mu\phi\nabla_\nu\phi - \delta^\mu_\nu\nabla_\lambda\phi\nabla^\lambda\phi\,, \\
\label{S}
\mathcal{S} &\equiv \mathcal{S}^{\mu\nu}_{\mu\nu} = \phi^2 R-6\phi\square\phi \, .
\end{align}
One can see that Eqs.~\eqref{Yamabe} and~ \eqref{S} coincide, up to an overall factor, if and only if $c=0$. A natural generalization of Eq.~\eqref{Stensor} including the missing piece of the Yamabe operator in Eq.~\eqref{S} can be obtained by shifting $S^{\mu\nu}_{\lambda\rho}$ in field space according to
\begin{equation}\label{Sigma}
\Sigma^{\mu\nu}_{\lambda\rho} = \frac{1}{\phi^2}\left(\mathcal{S}^{\mu\nu}_{\lambda\rho} + 2\nu\phi^4\,\delta^{\mu\nu}_{\lambda\rho} \right)\,,
\end{equation}
where $\delta^{\mu_1\ldots\mu_p}_{\nu_1\ldots\nu_p}=p!\,\delta^{\mu_1}_{[\nu_1}\dots\delta^{\mu_p}_{\nu_p]}$ is the generalized Kronecker delta of rank $p$. Written in this way, one can see that $\Sigma^{\mu\nu}_{\lambda\rho}$ has the same conformal weight as the Weyl tensor. Then, its trace gives
\begin{align}\label{Yamabe2}
\phi^2\Sigma^{\mu\nu}_{\mu\nu} &=  \phi^2 R - 6 \phi\Box\phi + 24\nu\phi^4 = 6\phi\tilde{\Delta}_2\phi
\end{align}
for $c=4\nu$. Namely, the full trace of the conformally covariant tensor $\Sigma^{\mu\nu}_{\lambda\rho}$ is equivalent, up to an overall factor, to the Yamabe operator in 4D.

Based on these considerations, we consider a conformally invariant scalar-tensor theory whose dynamics is described by the action principle 
\begin{align}\notag
I_{\phi,\rm conf} &= \frac{\zeta}{4}\int\diff{^4}x\sqrt{|g|}\,\delta^{\mu_1\ldots\mu_4}_{\nu_1\ldots\nu_4}\Sigma^{\nu_1\nu_2}_{\mu_1\mu_2}\Sigma^{\nu_3\nu_4}_{\mu_3\mu_4} \\
&= 96\zeta\nu\int\diff{^4}x\sqrt{|g|}\,\left[\frac{1}{12}\phi^2 R - \frac{1}{2}\phi\Box\phi + \nu\phi^4 + \frac{1}{96\nu}\,\left(E_4 + \nabla_\mu J^\mu\right)\right] \label{Lagphi}  \,,
\end{align}
where $\zeta$ is a dimensionless parameter and
\begin{align}
J^{\mu} &= 8 \left[\phi^{-1} G^\mu_\lambda\nabla^\lambda\phi+ \phi^{-2} \left(\nabla^{\mu} \phi \Box \phi -\nabla^{\lambda} \phi\nabla_{\lambda}\nabla^{\mu} \phi \right) + \phi^{-3} \nabla^{\mu} \phi \nabla^{\lambda} \phi \nabla_{\lambda} \phi \right]\,,
\end{align}
with $G_{\mu\nu}=R_{\mu\nu} - \frac{1}{2}g_{\mu\nu}R$ being the Einstein tensor. Equation~\eqref{Lagphi} reproduces, up-to-a boundary term, exactly the Yamabe operator of the conformally coupled scalar-tensor action in Eq.~\eqref{Ibulkconf2}. 

As previously shown, the Gauss-Bonnet term does not transform covariantly under Weyl rescalings [see Eq.~\eqref{weylvarGB}]. Nevertheless, since the left-hand side of the equation
\begin{align}\label{GBS}
    \frac{1}{4\phi^4}\delta^{\mu_1\ldots\mu_4}_{\nu_1\ldots\nu_4}S^{\nu_1\nu_2}_{\mu_1\mu_2}S^{\nu_3\nu_4}_{\mu_3\mu_4} = \frac{1}{4}\delta^{\mu_1\ldots\mu_4}_{\nu_1\ldots\nu_4}R^{\nu_1\nu_2}_{\mu_1\mu_2}R^{\nu_3\nu_4}_{\mu_3\mu_4} + \nabla_\mu J^\mu\,,
\end{align}
transforms covariantly under Weyl rescalings by construction, we conclude that the divergence of $J^\mu$ compensates the non-homogeneous piece of $E_4$ under conformal transformations. 

The field equations of the theory given by the initial action \eqref{ccscalar} can be obtained by performing stationary variations of the action~\eqref{Lagphi} with respect to the metric and scalar field, giving
\begin{subequations}\label{EOMLagphi}
\begin{align}\label{eomlagphig}
    T_{\mu\nu} &\equiv \nabla_\mu\phi\nabla_\nu\phi - \frac{1}{2}g_{\mu\nu}\nabla_\lambda\phi\nabla^\lambda\phi + \frac{1}{6}\left(g_{\mu\nu}\Box - \nabla_\mu\nabla_\nu + G_{\mu\nu} \right)\phi^2 - \nu\phi^4 g_{\mu\nu} = 0 \,, \\\label{eomlagphip}
   \mathcal{E} &\equiv\Box\phi - \frac{1}{6}\phi R - 4\nu\phi^3 = 0\,, 
\end{align}
\end{subequations}
respectively. By taking the trace of Eq.~\eqref{eomlagphig} and comparing it to Eq.~\eqref{eomlagphip}, one has that $T= g^{\mu\nu}T_{\mu\nu}=\phi\,\mathcal{E}$. Also, the $\Sigma$-tensor is related to the $T_{\mu\nu}$ of Eq.~\eqref{eomlagphig} through
\begin{equation}\label{SigmaT}
\phi^{2}\Sigma^\mu_\nu=6\left(T^\mu_\nu - \frac{1}{2}T\delta^\mu_\nu\right)\,.
\end{equation}
Notice that a constant scalar field configuration, say $\phi=\phi_0$, reduces the theory~\eqref{Lagphi} to Einstein-AdS gravity. This case corresponds to the Einstein frame of the Weyl symmetry. Indeed, in order for the action to be written in terms of the usual Newton's constant $G_N$ and the AdS radius $\ell$, the choice
\begin{equation}\label{couplingsEinstein}
    \nu \phi_{0}^{2} = \frac{1}{2\ell^2} \;\;\;\;\; \mbox{and} \;\;\;\;\;  \zeta = \frac{\ell^2}{64\pi G_{N}}\,,
\end{equation}
is made. At the level of the equations of motion, one can see that Eq.~\eqref{eomlagphig} becomes the usual Einstein field equation when fixing the Weyl gauge as in Eq.~\eqref{couplingsEinstein}. Also, Eq.~\eqref{eomlagphip} simply becomes the constraint that the Ricci scalar should be fixed in terms of the AdS radius, as is the case for Einstein spaces. Therefore, it is evident that the theory admits the full Einstein-AdS family as solutions for a constant scalar field.

At the level of the action, one can now check how the aforementioned Weyl gauge choice implies that the Lagrangian reduces to that of renormalized Einstein-AdS gravity. Indeed, for the values of the coupling constants in Eq.~\eqref{couplingsEinstein}, the action~\eqref{Ibulkconf2} can be cast into the MacDowell-Mansouri form given in Eq.~\eqref{IMM}. This is the exact form of the on-shell Einstein-Hilbert action with negative cosmological constant augmented by the Gauss-Bonnet term with a fixed coupling; the latter provides a natural counterterm for renormalizing the Euclidean on-shell action and the conserved charges for asymptotically locally Einstein-AdS solutions~\cite{Aros:1999id,Aros:1999kt,Miskovic:2009bm}. The previous discussion can be resumed in the following relation
\begin{equation}\label{MacdowellMansouri}
    I_{\phi_{0}, \rm conf} = I^{\rm (ren)}_{\rm EAdS} \,.
\end{equation}
Therefore, we conclude that the action~\eqref{Lagphi} has a well-posed Einstein limit defined through the choice of Eq.~\eqref{couplingsEinstein} for a constant scalar field.

In order to study the finiteness of the theory
~\eqref{Lagphi} when AlAdS spacetimes are considered, we perform the off-shell decomposition of the Weyl tensor \eqref{weyltensor} in terms of the $\Sigma$ tensor \eqref{SigmaT} and the Schouten tensor. Since the Einstein tensor appears explicitly in the definition of $T_{\mu \nu}$, we can equivalently write the Schouten as
\begin{equation}
S^{\mu}_{\nu} = \frac{1}{2} \left(G^{\mu}_{\nu} +\frac{1}{3}R \delta^{\mu}_{\nu}\right) \,.
\end{equation}
Taking into account the Eqs.~\eqref{eomlagphig} and~\eqref{eomlagphip}, one can replace the Einstein tensor and the Ricci scalar in terms of $T_{\mu \nu}$ and its corresponding trace. When the latter is replaced in the definition of the Weyl tensor \eqref{weyltensor}, results into the off-shell form of the Weyl, this is,
\begin{equation}\label{Weyldecomp}
    W^{\mu\nu}_{\alpha \beta} = \Sigma^{\mu\nu}_{\alpha\beta} - \frac{2}{\phi^2}\left(6T^{[\mu}_{[\alpha}\delta^{\nu]}_{\beta]} - T\delta^{\mu\nu}_{\alpha\beta} \right)\,.
\end{equation}
Since the field equations imply that $T_{\mu\nu}=0$, we conclude that the $\Sigma$ tensor coincides with the Weyl tensor and the on-shell action becomes
\begin{equation}
I_{\phi,\rm conf}\big|_{\rm on-shell} = \zeta\int\diff{^4}x\sqrt{|g|}\,W^{\mu\nu}_{\lambda\rho}W^{\lambda\rho}_{\mu\nu}\,.
\end{equation}
Therefore, the fully conformally coupled scalar-tensor theory is on-shell equivalent to CG, which is finite for any AlAdS solution~\cite{Grumiller:2013mxa}. A particular example of this fact has been recently shown in Ref.~\cite{Barrientos:2022yoz} for charged Taub-NUT-AdS and Eguchi-Hanson solutions in presence of conformally coupled scalar fields.

In analogy to the Einstein-AdS case discussed in Sec.~\ref{Sec:CG}, demanding exact local conformal invariance of the action under Weyl rescalings of both the metric and the scalar field, dictates the counterterms that render the action finite. Namely, the relation $I_{\phi,\rm conf}=I_{\phi}^{\rm (ren)}$ is valid and it leads to the renormalized conformally coupled scalar action, that reads
\begin{align}
I_{\phi}^{\rm (ren)}&=\frac{1}{384\nu}\int\diff{^4}x\sqrt{|g|}\,\delta^{\mu_1\ldots\mu_4}_{\nu_1\ldots\nu_4}\Sigma^{\nu_1\nu_2}_{\mu_1\mu_2}\Sigma^{\nu_3\nu_4}_{\mu_3\mu_4}= I_{\phi}+\frac{1}{96\nu}\int \diff{^4}x\sqrt{|g|} \left(E_4 + \nabla_\mu \tilde{J}^\mu\right) \,,
\label{Iphiren}
\end{align}
where
\begin{equation}
\tilde{J}^{\mu} = 8 \left[\phi^{-1} G^\mu_\lambda\nabla^\lambda\phi+ \phi^{-2} \left(\nabla^{\mu} \phi \Box \phi -\nabla^{\lambda} \phi\nabla_{\lambda}\nabla^{\mu} \phi \right) + \phi^{-3} \nabla^{\mu} \phi \nabla^{\lambda} \phi \nabla_{\lambda} \phi \right]-\tfrac{1}{2}\phi\nabla^{\mu}\phi \,
\end{equation}
and $I_{\phi}$ is defined in Eq.~\eqref{ccscalar}.
Hence, the last expression is conformally invariant for all configurations allowed by the solution space of the theory. 

We emphasize that in the theory of Eq.~\eqref{ccscalar}, or equivalently of Eq.~\eqref{Iphiren}, the scalar field cannot be gauged away to recover Einstein-AdS gravity~\eqref{couplingsEinstein} without changing the asymptotic behavior. Therefore, when fixing the AlAdS condition in a configuration with a non-trivial scalar field, the physical spacetime is chosen as the one in which said scalar field is present. Thus, the scalar is physical as it will contribute to the asymptotic charges of the configuration, and in the AdS/CFT context, to the holographic sources.

\section{Coupling to Einstein-AdS gravity\label{Sec:F2Sigma2}}

In this section, we extend the application of the Conformal Renormalization prescription, in the case where Einstein-AdS gravity couples to the non-minimally conformally coupled scalar-tensor field theory. In this case, the generic form of the action reads
\begin{equation}
I_{\phi \rm EAdS} = \frac{1}{16 \pi G_{N}} \int \diff{^4}x\sqrt{|g|} \left(R-2 \Lambda\right) + I_{\phi} \,.
\label{IphiEads}
\end{equation}
In the last two sections, we have shown that the cancellation of the divergences for AlAdS spacetimes amounts to the requirement that the on-shell action is invariant under Weyl rescalings of the bulk fields. However, the on-shell conformal completion of action in Eq.~\eqref{IphiEads} for any configuration of the solution space is highly non-trivial. Nevertheless, there are certain sectors of the theory that allow us to render the corresponding action conformally invariant. In order to do so, we consider the metric and the scalar sector separately. As shown in the last section, the latter can be supplemented by boundary terms which make it off-shell conformally invariant in four dimensions. Although this is finite for all possible solutions of such a theory, it is not expected to be true when other sectors are included. In a similar fashion, the pure-metric sector of the theory renders conformally invariant for Einstein-AdS spacetimes when augmented by the Gauss-Bonnet term with a fixed overall constant, as seen in Eq.~\eqref{IMM} and in Refs.~\cite{Maldacena:2011mk,Anastasiou:2016jix}. Nevertheless, we show that, provided a suitable asymptotic behavior of the scalar field in AdS, the action is finite without reference to intrinsic boundary counterterms. 

\subsection{Renormalization}

The dynamics of the theory we are interested in is dictated by an action principle that contains an Einstein-AdS sector written in a MacDowell-Mansouri form and the renormalized conformally coupled scalars, namely,  
\begin{equation}\label{IMMphi}
    I_{\rm MM\phi} = \frac{1}{4}\int\diff{^4}x\sqrt{|g|}\,\delta^{\mu_1\ldots\mu_4}_{\nu_1\ldots\nu_4}\left(\alpha \mathcal{F}^{\nu_1\nu_2}_{\mu_1\mu_2}\mathcal{F}^{\nu_3\nu_4}_{\mu_3\mu_4} - \zeta \Sigma^{\nu_1\nu_2}_{\mu_1\mu_2}\Sigma^{\nu_3\nu_4}_{\mu_3\mu_4} \right)\,,
\end{equation}
where $\alpha = \frac{\ell^2}{64 \pi G}$, while $\mathcal{F}^{\mu\nu}_{\lambda\rho}$ and $\Sigma^{\mu\nu}_{\lambda\rho}$ are defined in Eqs.~\eqref{WeylE} and~\eqref{Sigma}, respectively. The field equations for the metric and scalar field obtained from arbitrary variations of the action~\eqref{IMMphi} with respect those fields are
\begin{subequations}\label{eomMMphi}
\begin{align}\label{eomgMMphi}
    \mathcal{E}_{\mu\nu} &\equiv \alpha \left(G_{\mu\nu} - \frac{3}{\ell^2} g_{\mu\nu}\right)  - 12 \ell^{2}\zeta\nu  T_{\mu\nu} =0 \,, \\
    \label{eomphiMMphi}
    \mathcal{E} &\equiv \Box\phi - \frac{1}{6}\phi R - 4\nu\phi^3 = 0\,,
\end{align}    
\end{subequations}
respectively. It is worth mentioning that this theory admits Einstein-AdS spaces as solutions when the scalar field is constant. Indeed, the condition of Eq.~\eqref{couplingsEinstein} imposes that the solutions to the Eqs.~\eqref{eomMMphi} are Einstein manifolds, for which the action of Eq.~\eqref{IMMphi} vanishes identically. In particular, this theory admits global AdS space as the ground state when the scalar field is constant. 

Having partially conformally completed the theory, we study under which conditions this can be rendered finite. Following the prescription introduced in the last section, we can rewrite the action in terms of the Weyl squared term, that is finite for any AlAdS spacetimes. In a similar fashion, we have introduced an alternative decomposition of the Weyl tensor in terms of the $\Sigma$ tensor~\eqref{Weyldecomp}.
Additionally, since the trace of the stress-energy tensor $T_{\mu\nu}$ vanishes on shell, then Eq.~\eqref{eomgMMphi} constrains the space of solutions to possess negative constant Ricci scalar, i.e.
\begin{equation}
    R = -\frac{12}{\ell^2}\,.
\end{equation}
This simplifies the $X$ tensor in Eq.~\eqref{Xtensor} as it is now depends explicitly on the traceless Ricci tensor. Replacing Eqs.~\eqref{weylfdecomposition} and~\eqref{Weyldecomp} into the action $I_{\rm MM\phi}$ and taking into account the EOM, we get that
\begin{equation}
I_{\rm MM\phi}\Big|_{\rm on-shell} = \int\diff{^4}x\sqrt{|g|} \left[\left(\alpha -\zeta\right) W^{\alpha \beta}_{\mu \nu} W^{\mu \nu}_{\alpha \beta} + 2 \alpha \left( \frac{\alpha}{4 \ell^{2} \zeta \nu^{2} \phi^{4}}-1\right) H^{\mu}_{\nu} H^{\nu}_{\mu}\right] \,. \label{IMMphios}
\end{equation}
This on-shell action matches the CG action for Einstein spaces $\left(H^{\mu}_{\nu} = 0\right)$ or, equivalently, for stealth solutions: a nontrivial scalar field with vanishing stress-energy tensor~\cite{Ayon-Beato:2004nzi,Ayon-Beato:2005yoq,Hassaine:2006gz,Ayon-Beato:2013bsa}. Therefore, in those cases the theory is finite for AlAdS Einstein spacetimes. Indeed, if $\alpha=\zeta$, the action vanishes identically for Einstein solutions, providing a different sort of criticality in scalar-tennsor theories. On the other hand, non-Einstein solutions can also provide a finite on-shell action if and only if the falloff of the non-Weyl contribution of the last expression is sufficiently fast. A characteristic example of the latter is given below.

\subsection{Applications: The MSTZ black hole}

In absence of the cosmological constant and the quartic potential, the first black hole solution to the field equations~\eqref{eomMMphi} was found in Refs.~\cite{Bocharova:1970skc,Bekenstein:1974sf}. However, in that case the scalar field diverges at the horizon. In order to circumvent this problem, the authors of Refs.~\cite{Martinez:2002ru,Martinez:2005di} showed that the addition of these two terms allows for the scalar-field singularity to lie behind the horizon.\footnote{For different solutions in this theory and modifications thereof, see Refs.~\cite{deHaro:2006ymc,Anabalon:2009qt,Charmousis:2009cm,Anabalon:2012tu,Bardoux:2013swa,Astorino:2013sfa,Giribet:2014bva,Astorino:2014mda,Cardenas:2014kaa,Cisterna:2018hzf,Aviles:2018vnf,Cisterna:2019uek,Caceres:2020myr,Cisterna:2021xxq,Karakasis:2021rpn,Barrientos:2022avi}} This is known as the Martinez-Staforelli-Troncoso-Zanelli  (MSTZ) black hole. In that case, the authors considered a line element that remains locally invariant under the action of the isometry groups $\mbox{SO}(3)\times\mathbb{R}$, $\mbox{SO}(1,2)\times\mathbb{R}$, and $\mbox{ISO}(2)\times\mathbb{R}$. These conditions yield
\begin{equation}\label{ansatz}
    \diff{s^2} = -f(r)\diff{t^2} + \frac{\diff{r^2}}{f(r)} + r^2\diff{\Sigma_{(k)}^2} \,.
\end{equation}
Here, $\diff{\Sigma_{(k)}^2}$ is the line element of a two-dimensional base manifold of constant curvature $k$, describing locally $\mathbb{S}^2$, $\mathbb{T}^2$, and $\mathbb{H}^2$ transverse sections for $k=1,0,-1$, respectively. The solution is then given by~\cite{Martinez:2002ru,Martinez:2005di}
\begin{equation}\label{MTZ}
    f(r) = k\left(1+\frac{\mu\,G}{r}\right)^2 + \frac{r^2}{\ell^2} \;\;\;\;\; \mbox{and} \;\;\;\;\; \phi(r) = \frac{1}{\ell}\sqrt{\frac{1}{2\nu}}\,\frac{\mu\,G}{r+\mu\,G}\,,
\end{equation}
where $\mu$ is an integration constant, $\nu>0$, and the condition on the parameters 
\begin{equation}
    \zeta = \frac{\ell^2}{64\pi G}\,,
\end{equation}
must be met. Indeed, we can fix $\zeta=\tfrac{1}{96\nu}$ without loss of generality in order to obtain exactly the same normalization as in Refs.~\cite{Martinez:2002ru,Martinez:2005di}. This solution has a curvature singularity at $r=0$. The cosmic censorship conjecture requires the existence of a horizon at $r=r_h$, defined by the positive real roots of the polynomial $f(r_h)=0$. This condition demands that $k=-1$. In order for this solution to describe a black hole, the topology of the transverse section should be $\mathbb{H}^2/\Gamma$ where $\Gamma$ is a subgroup of $SO(2,1)$, such that the hypersurfaces of constant $t-r$ have a finite
area. Focusing on the case $\mu>0$, the solution has only one horizon and it is given by
\begin{equation}
    r_+ = \frac{\ell}{2}\left(1+\sqrt{1+\frac{4\mu G}{\ell}} \right)\,.
\end{equation}

To first order in the saddle-point approximation, we can obtain the partition function $\mathcal{Z}$ through the relation $\ln \mathcal{Z} \approx - I_E$, where $I_E$ is the Euclidean on-shell action. To compute the latter, we perform the analytic continuation to Euclidean time, that is, $t\to-i\tau$. The absence of conical singularities at the horizon implies that $\tau\sim\tau+\beta$, where $\beta$ is the period of the Euclidean time which is identified as the inverse of the Hawking temperature $T_H$. For the solution~\eqref{MTZ}, we find
\begin{equation}
T_H = \beta^{-1} = \frac{2r_+-\ell}{2\pi\ell^2}\,.
\end{equation}
Then, evaluating the Euclidean on-shell action~\eqref{IMMphi} on the solution~\eqref{MTZ} we obtain
\begin{equation}
    I_E = - \frac{\beta\omega_{(k)}\left(\ell^2-r_+\ell+r_+^2 \right)}{8\pi G\ell}\,,
\end{equation}
where $\omega_{(k)}$ is the volume of the codimension-2 base manifold. Remarkably, the value of the partition function is finite without any reference to boundary counterterms, even though it corresponds to a non-stealth configuration. This is due to the fact that the falloff of the scalar field and the traceless Ricci tensor is of order $\mathcal{O} \left(r^{-1}\right)$ and $\mathcal{O} \left(r^{-4}\right)$, respectively, what makes the non-Weyl contribution in Eq.~\eqref{IMMphios} to be subdominant, inducing no divergences. The latter can be seen directly from the fact that
\begin{align}
\int\diff{^4x}\sqrt{|g|}\left( \frac{\alpha}{4 \ell^{2} \zeta \nu^{2} \phi^{4}}-1\right) H^{\mu}_{\nu} H^{\nu}_{\mu} \sim \mathcal{O}(r^{-1})\,,
\end{align}
when evaluated at the solution~\eqref{MTZ}. Thus, this prescription provides a natural definition of counterterms for scalar-tensor theories possessing an Einstein sector and conformally coupled scalar fields whenever the non-Weyl squared part of the on-shell Lagrangian~\eqref{IMMphios} have a falloff that is at least as fast as $\mathcal{O} \left(r^{-4}\right)$. 

Furthermore, notice that the on-shell form of the action $I_{\rm MM\phi}$ indicates that it vanishes identically for Einstein spacetimes when $\alpha=\zeta$ or, equivalently, $\nu=\frac{2\pi G}{3\ell^2}$. Indeed, this is exactly the point in parameter space where the solution~\eqref{MTZ} exists, although the configuration is non-Einstein. This theory is completely analogous to the Critical Gravity, introduced in Ref.~\cite{Lu:2011zk}, that is trivial for Einstein spacetimes. This means that $I_{\rm MM\phi}$ for the specific value of $\nu$, corresponds to the generalization of Critical Gravity to scalar-tensor theories of gravity.

\section{Conformal gravity and scalar fields\label{Sec:CGscalars}}

Another very interesting class of scalar-tensor theories with manifest conformal invariance is given by the CG action~\eqref{ICGaction} and the non-minimally coupled scalar field action $I_{\phi}$, that is,
\begin{equation}
I_{\rm CG\phi} = I_{\rm CG} + I_{\phi} \,.
\label{ICGphi}
\end{equation}
Unlike the Einstein-Hilbert action in the previous section, the CG action is conformally invariant and the only part that it should be conformally completed comes from the scalar sector. This is achieved by introducing the surface terms that allows us to bring $I_{\phi}$ in the from given in Eq.~\eqref{Iphiren}. Thus, both actions are fully conformally invariant and they can be cast into the form
\begin{equation}
I_{\rm CG\phi,conf}   \equiv \frac{1}{4}\int\diff{^4x}\sqrt{|g|}\,\delta^{\mu_1\ldots\mu_4}_{\nu_1\ldots\nu_4}\left(\alpha W^{\nu_1\nu_2}_{\mu_1\mu_2}W^{\nu_3\nu_4}_{\mu_3\mu_4} - \zeta\Sigma^{\nu_1\nu_2}_{\mu_1\mu_2}\Sigma^{\nu_3\nu_4}_{\mu_3\mu_4}\right)  \,.\label{Lagconformal}
\end{equation}
In fact, this is the most general conformally invariant scalar-tensor theory of gravity, constructed out of squares of a single conformally-covariant tensor in the presence of scalar fields.
The field equations are obtained by performing stationary variations of the action with respect to the metric and scalar field, giving
\begin{subequations}\label{eom}
\begin{align}\label{eomg}
    \mathcal{E}_{\mu\nu} &\equiv \alpha B_{\mu\nu} - 48\zeta\nu\, T_{\mu\nu} = 0\,, \\
    \label{eomp}
    \mathcal{E} &\equiv \Box\phi - \frac{1}{6}\phi R - 4\nu\phi^3 = 0\,, 
\end{align}
\end{subequations}
respectively. For a constant scalar field, these field equations admit Einstein spaces as solutions since they are Bach-flat and have vanishing stress-energy tensor.

\subsection{Renormalization}

In order to study the consequences of the conformal completion of $I_{\rm CG\phi}$ on its renormalization, we rewrite the action in terms of Weyl squared taking into account the decomposition given in Eq.~\eqref{Weyldecomp}. In this case, $I_{\rm CG\phi,conf}$ can be written as
\begin{equation}\label{LagSigmaT}
    I_{\rm CG\phi,conf} = \int\diff{^4x}\sqrt{|g|}\left[ \frac{(\alpha-\zeta)}{4}\delta^{\mu_1\ldots\mu_4}_{\nu_1\ldots\nu_4}W^{\nu_1\nu_2}_{\mu_1\mu_2}W^{\nu_3\nu_4}_{\mu_3\mu_4} + \frac{24\zeta}{\phi^4}\left(3 T_{\mu}^{\nu}T_{\nu}^{\mu}-T^2\right)\right]\,.
\end{equation}
This is just a rewriting of the action in Eq.~\eqref{Lagconformal}, which can be further simplified when going on-shell due to Eq.~\eqref{eomp} or, equivalently, $T=0$, giving
\begin{equation}\label{LagSigmaTon-shell}
I_{\rm CG\phi,conf}\Big|_{\rm on-shell} = \int\diff{^4x}\sqrt{|g|}\left[ \frac{(\alpha-\zeta)}{4}\delta^{\mu_1\ldots\mu_4}_{\nu_1\ldots\nu_4}W^{\nu_1\nu_2}_{\mu_1\mu_2}W^{\nu_3\nu_4}_{\mu_3\mu_4} + \frac{72\zeta}{\phi^4}T_{\mu}^{\nu}T_{\nu}^{\mu}\right]\,.
\end{equation}
Thus, the action matches the Weyl squared, acquiring its well-behaved asymptotics for any AlAdS spacetimes, when configurations corresponding to $T_{\mu\nu}=0$ are considered. Furthermore, as noticed in the last section, the addition of the counterterms that conformally complete $I_{\phi}$ lead to a finite action even for non-vanishing $T_{\mu \nu}$ configurations, as long as the fall-off of the term $\phi^{-4}T_{\mu}^{\nu}T_{\nu}^{\mu}$ is fast enough.

Furthermore, there exists a particular point in parameter space, i.e. $\alpha=\zeta$, where the action vanishes identically for all solutions of the theory for which $T_{\mu \nu}=0$. Thus, the Euclidean continuation of these solutions have vanishing on-shell action like the maximally-symmetric solution, and therefore they can be considered to be part of the same vacuum state of the theory. The latter condition does not imply necessarily that the scalar field has to be trivial, rather, it could represent a stealth configuration. This yields an extended notion of criticality since the Euclidean on-shell action vanishes for stealth solutions at this particular point in parameter space. This results is in complete analogy to Critical Gravity for the metric fields~\cite{Lu:2011zk} where the action and conserved charges vanish identically for Einstein spaces~\cite{Miskovic:2014zja,Anastasiou:2017rjf}. However, in both theories the vacuum sector is determined by Einstein spacetimes. Here, on the other hand, criticality arises for stealth configurations that can or cannot be Einstein spacetimes.

\subsection{Applications: Stealth configurations over the Riegert Black hole}

Let us study stealth configurations to check Conformal Renormalization explicitly. Replacing Eq.~\eqref{ansatz} into~\eqref{eom}, the following solution is found
\begin{subequations}\label{solution}
\begin{align}
    f(r) &= k + \frac{6mG}{r_0} - \frac{2}{r_0}\left(k+\frac{3mG}{r_0} \right)r - \frac{2mG}{r} - \frac{\lambda r^2}{3}\,,\\
    \phi(r) &= \frac{1}{r-r_0}\sqrt{-\frac{k+\frac{2mG}{r_0}+\frac{\lambda r_0^2}{3}}{2\nu}}\,,
\end{align}
\end{subequations}
where $m$, $\lambda$, and $r_0$ are integration constants. This solution exists only for $\nu\neq0$ and $r_0\neq0$ and, to the best of our knowledge, it was found first in~\cite{Brihaye:2009ef}. Indeed, even though the scalar field is nontrivial, it has a vanishing stress-energy tensor. Thus, we conclude that this solution represents a stealth scalar field~\cite{Ayon-Beato:2004nzi,Ayon-Beato:2005yoq,Hassaine:2006gz,Ayon-Beato:2013bsa} over the Riegert metric~\cite{Riegert:1984zz}; the latter represents the most general static and spherically symmetric Bach-flat spacetime that solves the field equations of conformal gravity. Moreover, this solution is continuously connected to the topological Schwarzschild black hole AdS when $r_0\to\infty$, where the scalar field becomes constant, namely, $\phi=\sqrt{-\frac{\lambda}{6\nu}}$.   

The solution is endowed with a curvature singularity at $r=0$ as it can be seen from its Ricci scalar, that is,
\begin{equation}
    R = 4\lambda + \frac{12(3mG+kr_0)}{r_0^2\,r} - \frac{12mG}{r_0\,r^2}\,.
\end{equation}
The singularity is dressed by a single horizon defined as the real root of the cubic polynomial $f(r_h)=0$. We do not provide its explicit expression because it is cumbersome and not very illuminating. However, in order for a black hole horizon to exist, the condition $r_h>0$ must be met. Then, focusing on the case $k=1$ for the sake of simplicity, we identify two possible cases: (i)~$r_0>0$ or (ii)~$r_0<0$. In both cases, we find that $m>0$. In the first one, there exists a pole in the scalar field at $r=r_0$. Then, demanding that the scalar pole lies behind of the horizon, we find $0<r_0<6mG<r_h$ and
\begin{align}
    \frac{(r_0-6mG)(r_0+3mG)^2}{r_0(3 mGr_0)^2 }&<\Lambda <0 &\lor& & \Lambda &=\frac{(r_0-6mG)(r_0+3mG)^2}{r_0(3 mGr_0)^2}\,.
\end{align}
On the other hand, if $r_0<0$, the scalar field is regular for $r\in\mathbb{R}_{>0}$. Then, there is no need to demand $r_h>r_0$. Thus, we find 
\begin{align}
    \bigg(r_0&\leq -2mG& &\land& \Lambda &\leq-\frac{3(r_0+2mG)}{r_0^3}\bigg)& &\lor& \bigg(r_0&>-2mG & &\land&  \Lambda &<0\bigg)\,.
\end{align}
These conditions guarantee the existence of a black hole solution with a regular scalar field outside the horizon.  

The black hole temperature of the solution~\eqref{solution} is given by
\begin{align}\label{beta1}
 T_H = \frac{(r_h-r_0)\left[k(3r_h-r_0) - \lambda r_h^2(r_h-r_0)\right]}{4\pi r_h \left[ 3r_h(r_h-r_0)  +r_0^2\right]} \,,
\end{align}
which provides the inverse of the Euclidean's time period of this configuration. As we anticipated, conformal invariance of the action functional~\eqref{Lagconformal} renders the Euclidean on-shell action finite for stealth configurations. This can be seen explicitly by evaluating the Eq.~\eqref{Lagconformal} in the solution~\eqref{solution}, giving
\begin{align}
    I_E &= -\frac{16(\alpha-\zeta)\beta\omega_{(k)}m^2G^2}{r_h^3}\left[1-\frac{3r_h}{r_0}\left(1-\frac{r_h}{r_0}\right) \right]\,.
\end{align}
Notice that, even though the scalar field does not backreact, its presence modifies the Euclidean on-shell action in a nontrivial way. Then, we can unveil whether the thermodynamic system develops a Hawking-Page phase transition when $\alpha\neq\zeta$. This will be studied in a forthcoming work.

Additionally, there exist another asymptotically AdS black hole which is not continuously connected to the solution presented in Eq.~\eqref{solution}. These configurations were first reported in Ref.~\cite{Herrera:2017ztd}, where their conserved charges were obtained through the Abbott-Deser-Tekin formalism~\cite{Abbott:1981ff,Deser:2002rt,Deser:2002jk,Deser:2003vh}. In the coordinates we are using here, the asymptotically AdS solution reported in~\cite{Herrera:2017ztd} is given by
\begin{subequations}\label{solutionsyerko}
\begin{align}
    f(r) &= k + 12\nu\phi_0 -br - \frac{\lambda r^2}{3}\,, \\
    \phi(r) &= \frac{\sqrt{\phi_0}}{r}\,,
\end{align}
\end{subequations}
where $\phi_0$, $b$, and $\lambda$ are integration constants subject to the condition $(\zeta-4\alpha)(k+6\nu\phi_0)=0$. Although this solution is not continuously connected to~\eqref{solution}, it represents a stealth scalar field configuration as well. This solution possesses a curvature singularity at $r=0$ which can be dressed by a horizon located at the locus $r=r_h$ defined through the polynomial $f(r_h)=0$. In the Euclidean section, the absence of conical singularities demands that the period of the Euclidean time is
\begin{align}
    \beta &= \frac{4\pi r_h\ell^2}{r_h^2+k\ell^2}\,,
\end{align}
from which one can read the Hawking temperature. The conformally renormalized Euclidean on-shell action~\eqref{Lagconformal} for the solution in Eq.~\eqref{solutionsyerko} is then given by
\begin{align}
    I_E = -\frac{16\beta\omega_{(k)}k^2(\alpha-\zeta)}{3r_h}\,.
\end{align}
Thus, we conclude that Conformal Renormalization provides a finite value of the Euclidean on-shell action for the solution~\cite{Herrera:2017ztd} as well. We will study thermodynamics of this system in the extended phase space formalism in a forthcoming work.  


\section{Discussion\label{Sec:discussion}}

We studied how the renormalization of non-minimally coupled scalar-tensor gravity theories is dictated by restoring the on-shell conformal symmetry in the bulk. We considered the case of a conformally-coupled scalar field with a quartic potential, whose theory produces a boundary term when performing a Weyl transformation. Then, the conformal symmetry is restored by writing the kinetic term of the scalar in a non-canonical way, such that the action could be written in terms of the Yamabe operator. The resulting action was Weyl invariant except for the constant scalar configuration, where the conformal transformation became singular as evidenced by the fact that the action was divergent for AlAdS spacetimes. The restoration of the local rescaling symmetry was achieved by defining the $\Sigma^{\mu \nu}_{\rho \lambda}$ tensor as given in Eq.~\eqref{Sigma}, in terms of the metric and scalar degrees of freedom, such that it is conformally covariant and it has the same conformal weight as the Weyl tensor. Then, $\Sigma$-squared is a local conformal invariant of the theory, which can be used to define the action such that the theory is both renormalized and has full conformal invariance restored ---such that the Weyl variation vanishes exactly and there are no singular points in the transformation. The resulting action was shown to be on-shell renormalized, such that any field configuration that satisfies the EOM of the theory has a finite action when considering AlAdS spacetimes with weakened AdS asymptotics. Additionally, we note that the scalar field cannot be gauged away without modifying the boundary behavior away from the AlAdS condition, and thus it would change the physical state.

We also studied the theory which considers renormalized Einstein-AdS gravity written in MacDowell-Mansouri form in the presence of the renormalized conformally coupled scalar field. In that case, the theory does not have conformal invariance due to the Einstein-Hilbert sector. However, for Einstein spaces and constant scalar field, it becomes on-shell conformal invariant. In this case, the theory is renormalized for Einstein spacetimes as the full action becomes proportional to Weyl squared. Other configurations which also have finite action are those with AlAdS metrics such that their non-Einstein degrees of freedom, encoded in the traceless Ricci tensor, vanish sufficiently fast towards the conformal boundary. The MSTZ black hole~\cite{Martinez:2002ru,Martinez:2005di} belongs to this category and the value of the on-shell action and the corresponding black hole temperature were also computed.

Then, the same analysis was performed in conformal gravity plus renormalized conformally coupled scalar fields, which corresponds to the most general locally conformal invariant action constructed from quadratic antisymmetric contractions of Weyl-covariant tensors. Indeed, we showed explicitly that this theory is renormalized for Bach-flat spacetimes, which by virtue of the EOM are also stealth configurations. For this type of spacetimes, the action is renormalized for AlAdS spacetimes as it is proportional to the CG action. As a particular example, stealth configurations over the Riegert metric were considered and their Euclidean on-shell action was shown to be finite.

In both theories, there are interesting points in parameter space where the action can be rendered trivial for certain types of metric and scalar configurations. In particular, for the theory that includes conformal gravity, the action vanishes at the critical point of $\alpha=\zeta$ for Bach-flat configurations. Analogously, the theory that includes an Einstein-AdS sector has a trivial action at the critical point of $\alpha=\zeta$ for Einstein spaces, which correspond to solutions to the EOM with a constant value of the scalar field.

The fact that the action is rendered trivial implies that all asymptotic charges vanish identically, as well as the thermodynamic potential which is proportional to the on-shell action in the Euclidean section. This suggests a novel notion of criticality in scalar-tensor theories, which is different from the standard definition formulated perturbatively in terms of a decoupling of the massive modes of the metric from the spectrum~\cite{Lu:2011zk}. This notion of thermodynamic criticality corresponds to points in theory space where the ground state of the theory becomes enhanced, such that the maximally-symmetric configuration has the same free energy as an entire class of solutions. Thus, they form a moduli-space of vacuum configurations, which would admit spontaneous transitions between them at zero free energy cost. This idea was already discussed in the case of pure gravity theories, for Einstein-AdS gravity in both 4D and 6D in Refs.~\cite{Anastasiou:2016jix,Anastasiou:2017rjf,Anastasiou:2021tlv}. Although this point is very interesting, its full analysis requires a careful study of the thermodynamics. The possible phase transitions, along with the holographic analysis of the renormalized actions, will be studied in a follow-up paper.

\begin{acknowledgments}

We thank to Eloy Ay\'on-Beato, Nicol\'as C\'aceres, Daniel Flores-Alfonso, Alberto G\"uijosa, Rodrigo Olea and Julio Oliva for insightful comments and discussions. The work of GA is funded by ANID, Convocatoria Nacional Subvenci\'on a Instalaci\'on en la Academia Convocatoria A\~no 2021, Folio SA77210007. The work of IJA is funded by ANID FONDECYT grants No.~11230419 and~1231133. IJA also acknowledges funding by ANID, REC Convocatoria Nacional Subvenci\'on a Instalaci\'on en la Academia Convocatoria A\~no 2020, Folio PAI77200097. MBB is partially supported by Becas de Magíster UNAP and by ANID through FONDECYT grants No~11180894 and~11200025. The work of CC is partially supported by ANID through FONDECYT grants No.~11200025, 1210500, and~1230112. NM was partially supported by ANID through FONDECYT grants No.~11180894 and~1231133.

\end{acknowledgments}

\bibliography{References.bib}

\end{document}